\documentclass[prl,aps,twocolumn,showpacs,groupedaddress]{revtex4}
\usepackage{epsfig,amssymb,amsmath,amsfonts,bbm}
\usepackage{graphicx}

\def\be{\begin{eqnarray}}
\def\ee{\end{eqnarray}}
\def\Tr{\mbox{Tr}}
\newcommand{\rr}{\mathbbm{R}}
\def\Det{\mbox{Det}}

\begin{document}

\title{Teleportation-Induced Correlated Quantum Channels}

\author{F. Caruso$^{1,2,3}$, V. Giovannetti$^{3}$, and G. M. Palma$^{4}$}

\affiliation{$^{1}$Institute for Mathematical Sciences, 53
Prince's Gate, Imperial College, London, SW7 2PG, UK}
\affiliation{$^{2}$QOLS, The Blackett Laboratory, Imperial
College, London, Prince Consort Road, SW7 2BW, UK}
\affiliation{$^3$NEST-CNR-INFM \& Scuola Normale Superiore, Piazza
dei Cavalieri 7, I-56126, Pisa, Italy} \affiliation{$^4$NEST-CNR -
INFM \& Dipartimento di Scienze Fisiche ed Astronomiche,
Universita' degli studi di Palermo, via Archirafi 36, Palermo,
I-90123, Italy}

\begin{abstract}

Quantum teleportation of a $n$-qubit state performed using as
entangled resource a general bipartite state of $2n$ qubits
instead of $n$ Bell states is equivalent to a correlated Pauli
channel. This provides a new characterization of such channels in
terms of many-body correlation functions of the teleporting media.
Our model is then generalized to the Continuous Variable case. We
show that this new representation provides a relatively simple
method for determining whether a correlated quantum channel is
able to reliably convey quantum messages by studying the
entanglement properties of the teleportation mediating system.

\end{abstract}

\date{\today}


\maketitle

Starting from the seminal works of~Ref.~\cite{MPV} an increasing
attention has been devoted to the study of the so called memory
or, more precisely, Correlated Quantum (CQ) communication
channels both at theoretical~\cite{KW,BM,VJP} and experimental level~\cite{EXP}. Differently from the standard
{memoryless} configurations,  a CQ channel is characterized by the
presence of correlated noise sources which operate jointly on
otherwise independently transmitted messages ({\em channel uses}
in jargon). Recently these communication lines  were
put in correspondence with the physics of many-body quantum
systems~\cite{GM,PV,RGM}, by representing their noise effects in
terms of unitary couplings connecting  the transmitted signals
with a many-body correlated quantum environment. In particular,
Ref.~\cite{PV}  pointed out the possibility of relating  the
capacity of CQ channels with environmental critical phenomena. In
this paper we introduce an alternative  approach to the
representation of CQ channels based on the teleportation
representation of  the (memoryless) depolarizing channels given by
Bowen and Bose~\cite{BB} and on its continuous variable
generalization~\cite{Ban}. This provides an alternative
characterization of  the noise correlations of the channel in
terms of many-body correlation functions of the {\em medium} that
is employed in the teleportation with respect to previously
analyzed models \cite{PV,GM,RGM}. Furthermore the  representation
introduced here provides a relatively simple method for
determining how reliable a CQ channel is to transmit quantum
information by analyzing the entanglement properties of the
physical system which mediates the teleportation. We start
presenting the model for finite dimensional system postponing the
continuous variable generalization to the second part of the
paper.

\begin{figure}[t!]
\centering
\includegraphics[width=0.37\textwidth,angle=0]{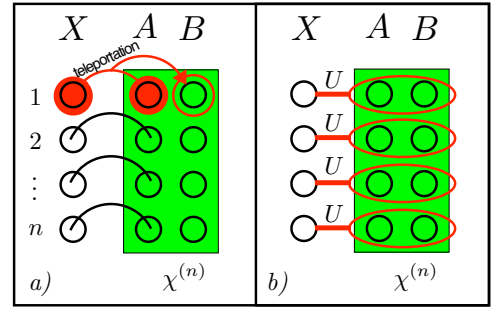}
\caption{Part {\em a)}:  Alice transfers messages encoded into the
input qubits $X$ into the $B$ part of the multi-particle medium
$\chi^{(n)}$ (green box) by teleportation (i.e. performing Bell
measurements on $XA$  and communicating to Bob the local
operations he has to perform on $B$). Part {\em b)}: both the
input and output states are written in $X$ while $\chi^{(n)}$
plays the role of an external (possibly) correlated environment --
see Eq.~(\ref{model1}). In both scenarios the resulting mapping is
given by Eq.~(\ref{due}).} \label{fig1}
 \end{figure}

\paragraph{Qubit CQ channels:--}

Consider $n$ {independent} quantum teleportation events~\cite{BEN}
in which the entangled resource initially shared by the
communicating parties (Alice and Bob) is not the correct Bell
state $|\Psi_{0}\rangle^{\otimes n}\equiv [(|00\rangle +
|11\rangle)/\sqrt{2}]^{\otimes n}$ but a bipartite  (not
necessarily pure) $2n$-qubit state $\chi^{(n)}$ -- the bipartition
being such that Alice has access on the first $n$ qubits $A$ ,
while Bob on the second $n$ qubits $B$. For $n=1$ it has been
shown~\cite{BB} that the transferring of a generic qubit state
$\rho$ from Alice to Bob can be seen as the result of the
depolarizing (or Pauli) channel $\Lambda^{(1)}(\rho) =
\sum_{k=0}^{3} p_{k} \sigma_{k}  \rho \; \sigma_{k}$, where
$\sigma_{k}$ is the $k$-th Pauli operator, and where $p_{k}$ are
the probabilities
 $p_{k} := \mbox{Tr} [ E_{k} \; \chi^{(n)}]$ with
$E_k := |\Psi_{k} \rangle\langle \Psi_{k} |= (\sigma_k \otimes
\openone) E_0 (\sigma_k  \otimes \openone) $ being the projectors
on the Bell basis  $|\Psi_{k}\rangle$ -- see Fig.~\ref{fig1} {\em
a)}. By the linearity this can be generalized to the $n>1$ case
yielding
\begin{eqnarray}
\Lambda^{(n)}(\rho) = \sum_{{\bold{k}}} p_{{\bold{k}}}^{(n)}\;
\sigma_{{\bold{k}}} \; \rho \;  \sigma_{{\bold{k}}}\;, \label{due}
\end{eqnarray}
where now $\rho$ is a generic input state of $n$ qubits, and for
${\bold{k}}:=(k_1,k_2, \cdots, k_n)$ one has
$\sigma_{{\bold{k}}}:=\sigma_{k_1} \otimes \cdots \otimes
\sigma_{k_n}$ and
\begin{eqnarray} \label{afaf}
p_{{\bold{k}}}^{(n)} := \mbox{Tr} [ E_{{\bold{k}}} \;  \chi^{(n)}] \;,
\end{eqnarray}
with $E_{{\bold{k}}} := E_{k_1} \otimes \cdots \otimes E_{k_n}$.
Equation~(\ref{due}) describes a $n$-qubit correlated Pauli
channel in which the rotations $\sigma_{\bf k}$ are randomly
applied to the system according to the joint probability
distribution $p_{\bold{k}}$ of finding $\chi^{(n)}$ into the state
$|\Psi_\bold{k}\rangle:= |\Psi_{{k_1}}\rangle\otimes \cdots
\otimes |\Psi_{{k_n}}\rangle$. Perfect teleportation is obtained
when $\chi^{(n)}$ coincides with $E_0^{\otimes n}$: in this case
$p_{\bold{k}}^{(n)} = \delta_{k_10} \cdots \delta_{k_n0}$ and
$\Lambda^{(n)}$ becomes the identity channel. Vice-versa total
depolarization is obtained, for instance, by using the complete
mixed state $\chi^{(n)} = \openone / 2^{2n}$ (here
$p_{{\bold{k}}}^{(n)} = 1/2^{2n}$ and $\Lambda^{(n)}(\rho) =
\openone/ 2^{n}$ for all $\rho$). It is finally worth pointing out
that the qubit CQ channels analyzed in Ref.~\cite{PV} are included
as the special case of our teleportation model, where all the $n$
shared couples of $\chi^{(n)}$  have vanishing projections on the
vectors $|\Psi_{1,2}\rangle$. Pauli channels form a convex set
which is closed under the super-operator composition. Specifically
concatenating two Pauli maps $\Lambda_{1}^{(n)}$ and
$\Lambda_{2}^{(n)}$ the result is still a Pauli channel
$\Lambda_{3}^{(n)}=\Lambda_{2}^{(n)} \circ \Lambda_{1}^{(n)} =
\Lambda_{1}^{(n)} \circ \Lambda_{2}^{(n)} $ whose
probabilities~(\ref{afaf}) are obtained by a double-stochastic
convolution of the previous ones (the order of the composition
being irrelevant). Vice-versa any Pauli channel
$\Lambda_{3}^{(n)}$ obtained by transforming the probabilities of
$\Lambda_{1}^{(n)}$ via  a double-stochastic transformation can be
expressed as $\Lambda_{2}^{(n)} \circ \Lambda_{1}^{(n)}$ (or
$\Lambda_{1}^{(n)} \circ \Lambda_{2}^{(n)}$) for some proper
choice of $\Lambda_{2}^{(n)}$. Expressed in terms of the
teleportation representation~(\ref{due})  these properties
translate into the following identities $\chi^{(n)}_3=
\Lambda_{1}^{(n)} (\chi^{(n)}_2)$ or  $\chi^{(n)}_3=
\Lambda_{2}^{(n)} (\chi^{(n)}_1)$ --- both choices yield the same
probabilities~(\ref{afaf}).

\paragraph{CJ isomorphism:--}

The teleportation representation  provides a {many to one}
correspondence among $2n$-qubit states $\chi^{(n)}$ and $n$-qubit
correlated Pauli channels. Indeed, since the off-diagonal terms of
the $\chi^{(n)}$s in the Bell basis do not influence the resulting
$p_{\bold{k}}^{(n)}$, we can associate more than one teleportation
mediator for each $\Lambda^{(n)}$. The simplest choice is of
course to use the Bell-diagonal state ${\chi}_{BD}^{(n)} :=
\sum_{\bold{k}} p_{\bold{k}}^{(n)} E_{\bold{k}}$. Interestingly
enough this also coincides with the Choi-Jamiolkowski (CJ) state
of $\Lambda^{(n)}$~\cite{CHOI}. We remind that the latter  is a
bipartite state associated with $\Lambda^{(n)}$ obtained by
applying the channel on half of a maximally entangled state
$|\Psi\rangle$ of the system and of an extra ancillary system,
i.e. $\left(\Lambda^{(n)} \otimes I \right)( | \Psi \rangle
\langle \Psi |) $ (here $I$ is the identity super-operator on the
ancilla). Its entropy and fidelity with $|\Psi\rangle$ coincide
respectively with channel entropy~\cite{ENTROPYCH} and the
entanglement fidelity~\cite{NIELSEN} of $\Lambda^{(n)}$ that were
used in Ref.~\cite{RGM} to characterize the correlations of CQ
channels. For the CQ channel of Eq.~(\ref{due}) we can take
$|\Psi\rangle:= |\Psi_0\rangle^{\otimes n}$ yielding the following
CJ state,
 \be
 \left(\Lambda^{(n)} \otimes I \right)( E_0^{\otimes n})
=\sum_{\bold{k}}
 p_{\bold{k}}^{(n)}\;
 E_{\bold{k}}  = \chi^{(n)}_{BD} \;,
 \label{choin}
 \ee
as anticipated. In Ref.~\cite{BDSW} it is shown that a
(memoryless) channel with a Bell-diagonal CJ state possesses an
useful symmetry (i.e., covariance property wrt Pauli rotations) in
such a way that having one use of that channel is both
mathematically and physically equivalent to having one copy of its
CJ state shared among the sender and the receiver of the message.
Therefore, by following the results of Ref.~\cite{BDSW} and
similar arguments as in Ref.~\cite{PV}, the quantum
capacity~\cite{QCAP} of any
family ${\cal L}:=\{ \Lambda^{(n)}: n
\in \mathbb{N}\}$ of CQ channels (\ref{due}) can be shown to obey
the following inequality
 \be\label{uupp}
 Q({\cal L}) \leq \lim_{n \rightarrow \infty} \frac{D_1(\chi^{(n)}_{BD}
 )}{n} \;,
 \ee
where, for each $n$, $\chi^{(n)}_{BD}$ is the CJ
state~(\ref{choin}) of $\Lambda^{(n)}$, and $D_{1}$ is the number
of e-bits we can extract from an {\em asymptotically large number
of copies} of $\chi^{(n)}_{BD}$ through one-way LOCC operations.
As a matter of fact, the inequality~(\ref{uupp}) holds also if we
replace $\chi_{BD}^{(n)}$ with a generic  $\chi^{(n)}$ that
provides a teleportation representation~(\ref{afaf}) for the maps
$\Lambda^{(n)}\in {\cal L}$ (this trivially follows by fact that
the direct teleportation through $\chi^{(n)}$ of e-bits is a
special example of one-way distillation). In particular, this
implies that $Q=0$ for  ${\cal L}$ admitting separable
teleportation media $\chi^{(n)}$. For forgetful channels~\cite{KW}
(i.e. for ${\cal L}$ having exponentially decreasing correlations,
e.g. Markov or memoryless channels) the upper bound in
Eq.~(\ref{uupp}) is tight. Thus, for these maps, by  exploiting
the hashing bound~\cite{BDSW,DW}, we can write
 \be
 Q({\cal L}) =\lim_{n \rightarrow \infty} \frac{D_1(\chi^{(n)}_{BD}
 )}{n}
 \geq
 1- \lim_{n \rightarrow \infty} \frac{S({\chi}_{BD}^{(n)})}{n}
 \label{hashQ}
 \;,
 \ee
with $S({\chi}_{BD}^{(n)})= -\sum_{\bold{k}} p_{\bold{k}}^{(n)}
\log_2 p_{\bold{k}}^{(n)}$ being the channel
entropy~\cite{ENTROPYCH} of $\Lambda^{(n)}$. This expression can
be further simplified when ${\chi}_{BD}^{(n)}$ is a {\em maximally
correlated state}~\cite{Rains2}. In this case the
inequality~(\ref{hashQ}) saturates and reduces to Eq.~(16) of
Ref.~\cite{PV} (see also Ref.~\cite{ABG}). It is finally worth
pointing out that for {\em all} (non necessarily forgetful) CQ
channels a upper bound~\cite{KW} can be obtained in terms of the
(regularized) coherent information $J(\rho, \Lambda^{(n)}) : =
S(\Lambda^{(n)}(\rho))  -S((\Lambda^{(n)} \otimes
I)(|\Psi_\rho\rangle\langle \Psi_\rho
 |) )$ of the channel, i.e.
\begin{eqnarray}
\label{eqnew1}
Q({\cal L}) \leq  \lim_{n \rightarrow \infty}\max_{\rho} J(\rho,
\Lambda^{(n)})/n\;,
\end{eqnarray}
(in this expression $|\Psi_\rho\rangle$ is a purification of
$\rho$).

\paragraph{Equivalence with other representations:--}

We now show how one can recast the channel~(\ref{due}) in terms of
an interaction with a many-body environment as in
Refs.~\cite{GM,PV,RGM}. To do so  we introduce the unitary
operator
\begin{eqnarray}
U^{(n)} := \sum_{\bold{k}} \sigma_{\bold{k}} \otimes E_{\bold{k}} = \otimes_{j=1}^n U_j  \;,
\end{eqnarray}
with $U_j = \sum_{k_j}  \sigma_{k_j} \otimes E_{k_j}$ the
transformation which rotates by $\sigma_{k_j}$ the $j$-th qubit of
the input ($X$ system of part  {\em b)} of Fig.~\ref{fig1}) while
projecting on the $k_j$-th Bell state the corresponding couples of
sites of teleporting medium $\chi^{(n)}$ ($AB$ system of the
figure). $U^{(n)}$ is thus a product of identical transformations
which act separately with each input qubit. The equivalence then
simply follows by
 the identity
\begin{eqnarray} \label{model1}
\Lambda^{(n)}(\rho) = \mbox{Tr}_{AB} \big[ U^{(n)} ( \rho \otimes \chi^{(n)} ) [U^{(n)}]^\dag \big]\;,
\end{eqnarray}
where the partial trace is performed over the teleporting medium.
In this representation  $\chi^{(n)}$ plays the role of the
many-body environment of the channel. This observation allows us
to take advantage of the results of Ref.~\cite{PV} to simplify the
analysis of the quantum capacity of our teleportation
representation. Specifically,  Plenio and Virmani showed that the
quantum capacity of the channel $\Lambda^{(n)}$ can be  expressed
as in the rhs term of Eq.~(\ref{hashQ}) if the  many-body
environmental state $\chi^{(n)}$ of Eq.~(\ref{model1}) admits a
decomposition in terms of translationally invariant matrix product
states~\cite{MPS} with finite bound dimensions.
Equation~(\ref{model1}) also shows that the weakly complementary
channel \cite{CGH}, mapping the initial system state to the final
environmental one, is given by $\tilde \Lambda^{(n)}(\rho) =
\mbox{Tr}_{X} \big[ U^{(n)} ( \rho \otimes \chi^{(n)} )
[U^{(n)}]^\dag \big]$. By taking $\chi^{(n)}$ in Bell diagonal
form it is easy to show that one has $\tilde \Lambda^{(n)}(\rho) =
\chi_{BD}^{(n)}$ for all $\rho$. Therefore, the CQ channel
$\Lambda^{(n)}$ in (\ref{model1}) is always weak-degradable
\cite{CGH}, i.e. ${\cal T} \circ \Lambda^{(n)} = \tilde
\Lambda^{(n)}$ with ${\cal T}= \tilde \Lambda^{(n)}$.

\paragraph{Examples:--}

We now analyze some examples of teleportation-induced CQ channels.

First, we consider the situation in which Alice and Bob share $n$
perfect Bell pairs $|\Psi_0\rangle^{\otimes n}$ but they
unfortunately loose the classical information about the order of
the particles (or equivalently a sequence of random swaps has been
applied to the qubits controlled by Alice). In this case  the
shared state between Alice and Bob, $\chi^{(n)}$, becomes a
$2n$-qubit mixture of all possible pairings. As found in
\cite{EFPPW}, the one-way distillable entanglement of $\chi^{(n)}$
is analytically calculated as $D_1(\chi^{(n)})=\sum_{j=0}^{n/2}
\frac{(2j+1)^2}{2^n(n+1)} {n+1 \choose n/2-j} \log(2j+1)$ (here
$n$ is assumed to be even). Replacing into Eq.~(\ref{uupp}) and
exploiting the fact that $D_1$ is exponentially decreasing in $n$
it is easy to find that, as expected, the quantum capacity of the
associated process ${\cal L}$ is vanishing, i.e. $Q=0$.

\begin{figure}[t!]
\centering
\includegraphics[width=0.4\textwidth,angle=0]{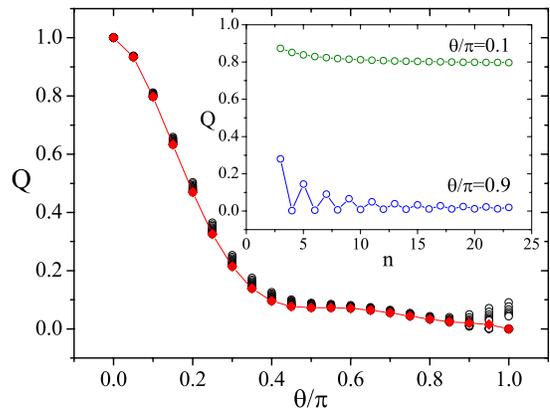}
\caption{Quantum capacity $Q({\cal L})$ of the family ${\cal L}$
of CQ channels as a function of the phase shift $\theta/\pi$ for
the quantum phase gate example (red diamond). For each value of
$\theta/\pi$, the values of $Q^{(n)}$, i.e.
$1-S({\chi}_{BD}^{(n)})/n$, are shown for $n=10,...,23$ from top
to bottom (empty circles). The ideal teleportation is recovered
for $\theta=0$, while $Q=0$ for $\theta/\pi=1$ because
$\chi^{(n)}$ becomes separable. In the inset, the fast convergence
($\lim_{n \rightarrow \infty}$) of $Q^{(n)}$ vs. $n$ is reported
for $\theta/\pi=0.1,0.9$. Similar behaviors are found for all
values $\theta/\pi \in [0,1]$.} \label{fig2}
 \end{figure}

As a second example, we consider the case in which the
teleportation mediator $\chi^{(n)}$ is obtained by applying (on
Alice side) a sequence of unitary  local transformations to $n$
Bell states $|\Psi_0\rangle$ which, as in the previous example,
are initially shared among Alice and Bob. Specifically we assume
that a quantum phase gate with a conditional phase shift $\theta$
is applied on Alice's $1^{st}$ and $2^{nd}$ qubits, later the same
gate is applied on her $2^{nd}$ and $3^{rd}$ one, and so on and so
forth for all $n$ qubits. One can easily verify that the resulting
state is a translationally invariant matrix product
state~\cite{MPS} which is maximally correlated (i.e. it has
vanishing projections on the vectors $|\Psi_{1,2}\rangle$).
Therefore  in this case the bound (\ref{hashQ}) is tight. In
particular, following~\cite{PV}, one can build an optimal code by
partitioning the channel's uses in blocks of \textit{live} and
\textit{spacer} qubits (in our case it is sufficient to have
spacer blocks of only two elements each). In Fig.~\ref{fig2} we
report the quantum capacity of this channel as a function of the
phase shift $\theta/\pi$ computed as in the rhs of
Eq.~(\ref{hashQ}). For $\theta=0$  one recovers the ideal
teleportation with maximum quantum capacity, while for
$\theta=\pi$ the state $\chi^{(n)}$ becomes separable and $Q$ is
exactly null.

\paragraph{CV systems:--}

We now generalize the teleportation representation of CQ channels
to  Continuous-Variable (CV) systems. Here we assume that Alice
and Bob share a bipartite $2n$-mode Bosonic state $\chi^{(n)}$
defined on an infinite Hilbert space ${\cal H}_A \otimes {\cal
H}_B$ which is organized in $n$ couples that are operated
independently during a CV teleportation stage. By following the
standard protocol~\cite{CVTP}, Alice performs the simultaneous
quantum measurement of position and momentum for the input system
in an arbitrary $n$-mode Bosonic state $\rho$ (to be teleported to
Bob) and her part of the bipartite system $\chi^{(n)}$. This is
described by a projection on the simultaneous eigenstate of
position-difference and momentum-sum. Then Alice informs Bob of
the measurement outcome and Bob applies a proper unitary
transformation $\hat{V}(z)$ on his part of $\chi^{(n)}$, with
$\hat{V}(z)$ being the multi-mode {Weyl} (displacement) operator
(i.e., $\hat{V}(z) := \exp [ i \hat{R} z]$ with $z:= (
x_1,x_2,\cdots, x_n, y_1 ,y_2, \cdots ,y_n)^T$ being a column
vector of ${\rr}^{2n}$ and $\hat{R} := ( \hat{Q}_1, \cdots,
\hat{Q}_n; \hat{P}_1, \cdots, \hat{P}_n)$, where $\{Q_i,P_i\}$ are
the canonical coordinates of $n$ modes~\cite{HOLEVO}). Building up
from the results of Ref.~\cite{Ban} for the $n=1$ case, the
resulting channel can now be described as  the following
generalized $n$-mode additive classical noise map~\cite{HW}
 \be
\Phi^{(n)}(\rho)=\int d z \ f^{(n)}(z) \ \hat{V}(z) \ \rho \
\hat{V}(z)^\dagger \;,
 \label{CVchannels}
 \ee
where $f^{(n)}(z)$ is the probability distribution
 \be
 f^{(n)}(z)= \Tr [E(z) \ \chi^{(n)}] \; , \label{n-modeP}
 \ee
with
 $E(z) = [\openone \otimes
\hat{V}(z)^\dagger] \ E_0^{\otimes n} \ [\openone \otimes
\hat{V}(z)]$, and $E_0$  being the (not normalized) CV version of
the Bell state~\cite{Ban}, i.e. the two-mode infinitely-squeezed
vacuum state. Equation~(\ref{n-modeP}) generalizes
Eq.~(\ref{afaf}) to the CV case. Here the CJ state becomes
 \be
\chi_{CJ}^{(n)} =\int d z \ f^{(n)}(z) \ E(z)
= \int d z  \ E(z)  \chi^{(n)} E(z)\;, \label{ChoiCV}
 \ee
which again can be used to replace $\chi^{(n)}$ in the
teleportation representation of $\Phi^{(n)}$. Following arguments
similar to those used in the qubit case, the quantum capacity
$Q({\cal F})$ of a family ${\cal F}:=\{ \Phi^{(n)}: n\in
\mathbb{N}\}$ can be bounded as in Eqs.~(\ref{uupp}), and again
the inequality saturates for forgetful channels. In the CV case an
upper bound for the distillable entanglement $D_1(\chi^{(n)})$ is
given by the logarithmic negativity ${\cal N}$, that is easily
computable in terms of the symplectic eigenvalues of the
Covariance Matrix (CM) of the CJ state~\cite{VW}. In particular,
let $\chi^{(n)}$ be a $2n$-mode Gaussian state and $E_0$ be the
density operator of a two-mode ($r \rightarrow \infty$)-squeezed
vacuum state and $\gamma_{E_0}$ its CM, i.e.
 $\gamma_{E_0}:=\tiny{\left[%
\begin{array}{cc}
  \Gamma_{-} & 0 \\
  0 & \Gamma_{+} \\
\end{array}%
\right]}$, where $\Gamma_{\pm}
:=\tiny{\left[%
\begin{array}{cc}
  a & b_{\pm} \\
  b_{\pm} & a \\
\end{array}%
\right]}$, $a= \nu \cosh(2 r)$, $b_{\pm}= \pm \nu \sinh(r)$, $\nu$
is the average photon number of the thermal state, and $r$ is the
squeezing parameter, which tends to infinity to have a CV Bell
state. Hence, since $E_0^{\otimes n}$ is $\gamma_{E_0}^{\oplus
n}$, we find
 \be
f^{(n)}(z) = \frac{2^{2n} \exp[-1/2 \ m^T (\gamma_{E_0}^{\oplus
n}+\gamma)^{-1} m]}{[\Det(\gamma_{E_0}^{\oplus
 n}+\gamma)]^{1/2}} \; , \;
 \ee
with $\gamma$ being the CM of  $\chi^{(n)}$ (assumed for the sake
of simplicity to have zero mean), and $m:=(0,z)$ being a real
vector of dimension $4n$. Then the logarithmic negativity ${\cal
N}$ is given by ${\cal N}=\tiny{\left\{
\begin{array}{ll}
-\sum_{k} \log \tilde{\nu}_k & \mbox{for $k: \tilde{\nu}_{k} < 1$}\\
 0 & \mbox{otherwise}
\end{array}
\right.} $,
 with $\tilde{\nu}_{k}$ the symplectic eigenvalues of the partial
transpose of the CM of $\chi_{CJ}^{(n)}$ \cite{VW}.

\paragraph{Conclusions.--}

We have presented a teleportation representation of correlated
Pauli channels and their CV generalizations which establishes a
formal connection between CQ maps and the physics of many-body
systems. Consequently, as in Refs.~\cite{PV,GM,RGM}, the CQ maps
can now be studied in terms of the properties  of the multi-qubit
state $\chi^{(n)}$ which mediates the teleportation. In
particular, in the qubit case,  by expanding the operators $E_k$
as a linear combination of the Pauli operators (e.g.,
$E_0=\sum_{k=0}^3 \sigma_k\otimes \sigma_k -2  \sigma_2\otimes
\sigma_2$) the probabilities $p_{\bold{k}}$ that characterize the
channel~(\ref{due}) can be expressed in terms of the $2n$-point
correlation functions  of the form $\mbox{Tr}[ \sigma_{\bold{k'}}
\otimes \sigma_{\bold{k''}}\;  \chi^{(n)}]$ (assuming $\chi^{(n)}$
to be the ground state of a many-body Hamiltonian $H$ the latter
can then be used to relate the correlations of
${\Lambda_\chi^{(n)}}$ to the physical system couplings).

F.C.\ and V.G. acknowledge the Quantum Information research
program of the Centro E. De Giorgi of SNS for financial support.
F.C. was supported also by the EU STREP project CORNER and he
would like to thank M.B. Plenio and S. Virmani for  discussions.

\end{document}